\documentclass[aps,pre,twocolumn,groupedaddress]{revtex4-2}

\usepackage{graphicx}
\usepackage{dcolumn}
\usepackage{bm}
\usepackage{hyperref}
\usepackage{subfigure}
\usepackage{xcolor}
\usepackage [english]{babel}
\usepackage{amsmath}
\usepackage{ulem}
\graphicspath{{Figures/}{./}} 

\newcommand{\efish}{E\babelhyphen{nobreak}FISH}
\newcommand{\changedtext}[1]{{\leavevmode#1}}
\newcommand{\removedtext}[1]{\ignorespaces} 

\begin{document}


\title{Measurement of the electric field distribution in streamer discharges}


\author{Yihao Guo$^1$, Anne Limburg$^1$, Jesse Laarman$^1$, Jannis Teunissen$^2$, Sander Nijdam$^1$}
\email[]{s.nijdam@tue.nl}
\affiliation{$^1$Department of Applied Physics and Science Education, Eindhoven University of Technology, 5600 MB Eindhoven, The Netherlands\\
$^2$Centrum Wiskunde \& Informatica (CWI), 1090 GB Amsterdam, The Netherlands}


\date{\today}

\begin{abstract}
Using electric field induced second harmonic generation (E-FISH), we performed direction-resolved absolute electric field measurements on single-channel streamer discharges in 70\,mbar \changedtext{(7\,kPa)} air with 0.2\,mm and 2\,ns resolutions. 
In order to obtain the absolute (local) electric field, we developed a deconvolution method taking into account the phase variations of E-FISH.
The acquired field distribution shows good agreement with the simulation results under the same conditions, in direction, magnitude and in shape. \changedtext{This is the first time that E-FISH is applied to streamers of this size (\textgreater 0.5\,cm radius), crossing a large gap.} Achieving these high resolution electric field measurements benefits further understanding of streamer discharges and enables future use of E-FISH on cylindrically symmetric (transient) electric field distributions.
\end{abstract}


\maketitle

Streamer discharges are fast propagating ionization fronts that appear as the precursor to lightning leaders and as sprites in nature \cite{Nijdam2020}. 
The electric field is the driving force behind streamers and determines their energy transfer and chemical activity \cite{Patnaik_2017}.
There are various studies focusing on electric field measurement of discharges using different diagnostics, including electric-field-induced coherent anti-Stokes Raman scattering (E-CARS)~\cite{vanderSchans_2017} and Stark spectroscopy~\cite{Cvetanović_2015}; while methods for measuring the electric field in streamers, which are highly transient, are limited, and if present, have very low temporal and spatial resolution.
Most recently, Dijcks~\textit{et al.}~\cite{Dijcks2023a} determined the electric field of single-channel streamers in pure nitrogen and synthetic air at pressures of 33\,mbar by using optical emission spectroscopy (OES).
However, this method has the disadvantages that it depends on light emission, cannot measure the field direction and has a rather low spatial and temporal resolution. The lack of suitable methods hinders further understanding of streamers and other transient discharges driven by the electric field.

To tackle these problems, a new technique called electric field induced second harmonic generation (E-FISH) has been introduced to the plasma community to measure the electric field of various kinds of plasmas~\cite{Dogariu2017,Chng2020a}. In this method, a high power laser beam non-linearly interacts with an electric field in a gas, generating second harmonics. The intensity of these second harmonics scales directly with the square of the electric field strength, and the polarization direction aligns with the field orientation.
Initially, this method was considered easy to implement and the measured signals straightforward to interpret.
However, it has been shown that the measured signals are strongly related to the laser beam profile, and depend heavily on the electric field profile and not just its integrated value due to phase variations along the laser beam~\cite{Chng2020}.
A solution has been proposed in \cite{PhysRevA.104.053511}, but uniformity of the electric field in one of the directions perpendicular to the laser beam is assumed, which is not a valid assumption for streamers. 

In this Letter, for the first time, we report detailed direction-resolved measurements of the electric field distribution in single-channel streamers in air by using \efish.
The electric field is restored from the \efish{} signals by applying a deconvolution method including all phase variations, where cylindrical symmetry is assumed.
Also, light emission from the discharge is captured.
Next to this, simulations on the electric field and the emission spectrum of the second positive system (SPS) of N$_2$ are obtained using a 2D axisymmetric drift-diffusion-reaction fluid model~\cite{Li2021} of the streamers under the same conditions. Li~\textit{et al.}~\cite{Li2021} have shown that this model is very reliable in simulating streamers in air; its calculated streamers closely match in velocities and diameters with experimental results.

The experimental results show a tremendous improvement in resolution compared to previous work. Thus far, such detailed electric field distributions of transient plasmas could only be obtained through simulations. Moreover, we reveal the electric field in areas with little to no light emission, which contains the most valuable information for streamer discharges. The magnitude, shape and direction of the field and its position relative to the light emission are in agreement with the simulations. This enables further understanding of streamers and empowers future use of E-FISH.

\begin{figure*}[!htb]
  \centering
  \includegraphics[width=0.9\textwidth]{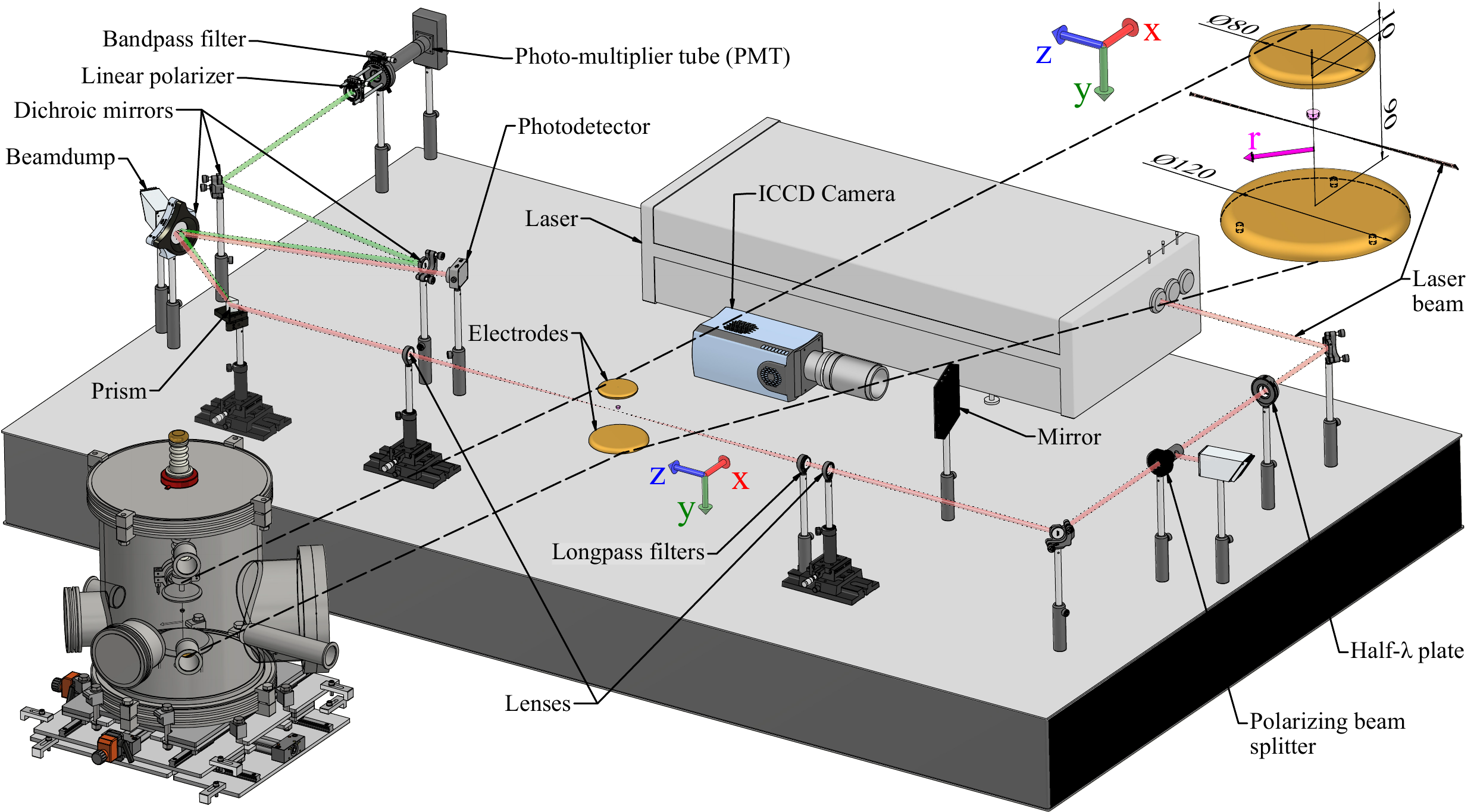}
  \caption{\small A render of the E-FISH setup.  The vacuum vessel is excluded from the main image for clarity and is instead shown on the bottom left. \changedtext{We use a coordinate system in which the streamer propagates in the $y$ direction, the laser beam is along the $z$ direction, and the radial coordinate is given by $r=\sqrt{x^2+z^2}$}. \removedtext{The coordinate system is indicated twice and} The detailed geometry of the electrodes including streamer, \changedtext{$r$-direction definition}, and the laser beam (dimensions in mm) is shown on the top right.}
  \label{fig:EfishSetup}
\end{figure*}


A schematic of the experimental setup is shown in Fig.~\ref{fig:EfishSetup}.
In short, a 100\,mJ Nd:YAG laser (EKSPLA SL234-10-G-SH) beam with a pulse width of 120\,ps at 1064\,nm is focused into the discharge area by an $f=500$\,mm lens. 
The beam waist and Rayleigh length $z_R$ are measured to be 0.13\,mm and 12\,mm, respectively, by using the knife edge method~\cite{Khosrofian:83}.
Two longpass filters remove any second harmonic light generated by the laser itself and by interaction between the laser beam and the upstream optics.
Second harmonic light is generated when the laser interacts with the electric field in the streamers that are produced inside a vessel.
The fundamental and second harmonic beams are collimated again by another $f=500$\,mm lens.
These two colinear beams are then fully separated by using a prism and three dichroic mirrors.
The fundamental beam intensity is measured by a photodetector (Thorlabs DET36A/M).
The second harmonic beam is directed to a photomultiplier (PMT, Hamamatsu H6779-04), with a bandpass filter attached in front of it to filter out any stray light.
To increase the signal-to-noise ratio, the laser beam is polarized parallel to the to-be-measured electric field direction by using a half-wave plate and a polarizer, because the second harmonic generation process has a higher efficiency under this configuration~\cite{Chng2020a}.
A polarizer is positioned in front of the PMT to measure the vector components of the electric field. 
\\ \indent To generate streamers, we use the setup that has been described in~\cite{Dijcks2023a}. 
The high voltage pulses are generated by a pulsed power circuit (Belhke HTS) with an amplitude of 9.5\,kV, a rise time of about 50\,ns, a voltage jitter of $\sim$1\,ns, and a duration of 400\,ns at a repetition rate of 60\,Hz.
The electrodes are protrusion-to-plane electrodes, where the high voltage electrode contains a 10\,mm protruding pin \changedtext{in the center} (1\,mm diameter, 60$^\circ$ tip angle, and 50\,\textmu m tip radius).
The pin-to-plate gap is 90\,mm. \changedtext{The laser beam is at the height of 5.3\,cm from the grounded bottom plate electrode.}
During the experiment, the vessel is continuously flushed with dry air with a flow rate of 2\,L/min and the pressure is fixed at 70\,mbar \changedtext{(7\,kPa)}.
The pulse repetition rate, the dedicated electrode geometry, and the reduced pressure ensure the high repetitiveness in both time and space of the single positive streamers, and thus the quality of the \efish~signals.
\\ \indent The moment the laser interacts with the streamer electric field, the discharge image is captured by an ICCD camera (Andor DH334T) with a gating time of 2\,ns. From these images, only a horizontal strip at the height of the laser beam is used.
Under the above mentioned conditions, the streamer propagation velocity is measured to be 3.5$\times 10^5$\,m/s.
The laser, the high voltage pulses, and the gating of the ICCD camera are synchronized by a digital delay generator.
\\ \indent In order to obtain the electric field at different positions in the discharge, the entire vessel is fixed on a translation stage such that it can move over the horizontal axis (coordinate $x$) perpendicular to the laser beam direction (coordinate $z$).
The second dimension is acquired by varying the delay between the high voltage pulse (and consequently streamer inception) and the laser trigger. 
This allows us to generate an image with time as one dimension and the horizontal cross-section of the streamer in the second dimension both for optical emission and electric field strength.
Because the single-channel streamers under investigation are roughly constant in shape and velocity when traversing the center of the electrode gap~\cite{Li2021}, the time axis is very similar to the streamer propagation direction axis (coordinate $y$).
For a single measurement of one field direction, a time range of $-$80 to 120 ns with a step size of 2\,ns and a spatial range of $-$15 to 15\,mm with a step size of 0.2\,mm is used, where \changedtext{$t=0$} \removedtext{zero} is defined as the moment when the electric field in the $y$-direction peaks, and \changedtext{$x=0$} the middle of the streamer in the $x$-direction \removedtext{, respectively}. For every delay and position, 40 laser shots are recorded, resulting in one measurement consisting of more than 400,000 laser shots (the grid size is coarser for the lower field region).
A full measurement, for both directions of the electric field, takes about 15 hours.

The signal we measure is $s(x,t) = \sqrt{I_{2\omega}/I_\omega^2}$, where $I_{2\omega}$ and $I_{\omega}$ represent the intensity of the second harmonic and fundamental beams, respectively.
This signal is the result of a line-of-sight integration along the laser propagating direction\changedtext{~\cite{Boyd2020,Chng2020}}:
\begin{equation}\label{eq:GBS_intensitySH}
    s(x,t) = C_\mathrm{cal} \left| \int_{-L}^L  \frac{e^{-i\Delta k z}}{1 + i \frac{z}{z_R}} E_\mathrm{ext}(x, z, t)  \, \mathrm{d}z \right | , 
\end{equation}
where $C_\mathrm{cal}$ is a calibration constant; $\changedtext{2}L$ is the interaction length of the laser beam and the electric field; $\Delta k$ is the wavevector mismatch between the fundamental and second harmonic wavelengths, \changedtext{$z_R$ is the Rayleigh length of the focused beam, } and $E_\mathrm{ext}$ the external, to-be-measured electric field\changedtext{ of which the $x$- and $y$-component can be measured}.
$\Delta k$ is 3.45 m$^{-1}$ for 70\,mbar air with a 1064\,nm input beam~\cite{Borzsonyi08}.
\changedtext{The complex denominator, $1+i\frac{z}{z_R}$, introduces an extra phase shift $\mathrm{tan^{-1}} \frac{z}{z_R}$ called Gouy phase, which originates from the focused Gaussian beam shape.} 
For an axisymmetric streamer, the electric field in the $y$-direction $E_y$ will also be axisymmetric, such that $E_\mathrm{ext}(x, z, \changedtext{t}) = E_y(r, \changedtext{t})$ where $r = \sqrt{x^2 + z^2}$.
We can also make use of the axisymmetry when measuring the field in the $x$-direction $E_x$, by expressing it as $E_x = \sin(\theta) E_r$, where $E_r$ is the axisymmetric radial component and $\sin(\theta) = x/r$.
  In this case, we thus have $E_\mathrm{ext}(x, z, \changedtext{t}) = x/r \, E_r(r,\changedtext{t})$.

We cannot use a standard inverse Abel transform to solve equation \eqref{eq:GBS_intensitySH}.
Instead, we approximate the integral by a weighted sum over samples $E_y(r_i)$ or $E_r(r_i)$, at a given set of radial coordinates.
Given a set of measurements $s(x_j)$, we can then solve an approximately linear system to obtain $E_x(r_i)$ or $E_y(r_i)$.
We include a small regularization parameter in this procedure to make the inversion unique and robust to noise in the measurements.
A detailed description of the inversion procedure is given in the supplementary material~\cite{Supplementary}.
Similarly, the optical emission is Abel-inverted.


To obtain the absolute field strength, a calibration measurement on a known field distribution is performed.
We designed rod-to-rod and rod-to-cylinder electrodes for the calibration of $E_y$ and $E_x$, respectively, which generate an electric field with a similar shape as the streamer.
First, the electrostatic field of these configurations is simulated in COMSOL Multiphysics. 
Then the field is forward-transformed by using the inverse of the algorithm described above to obtain the ``calculated \efish~ signals''.
The ratio between the calculated signals and the measured signals leads to the calibration constant $C_\mathrm{cal}$.

\begin{figure}[!htb]
  \centering
  \includegraphics[width=0.5\textwidth]{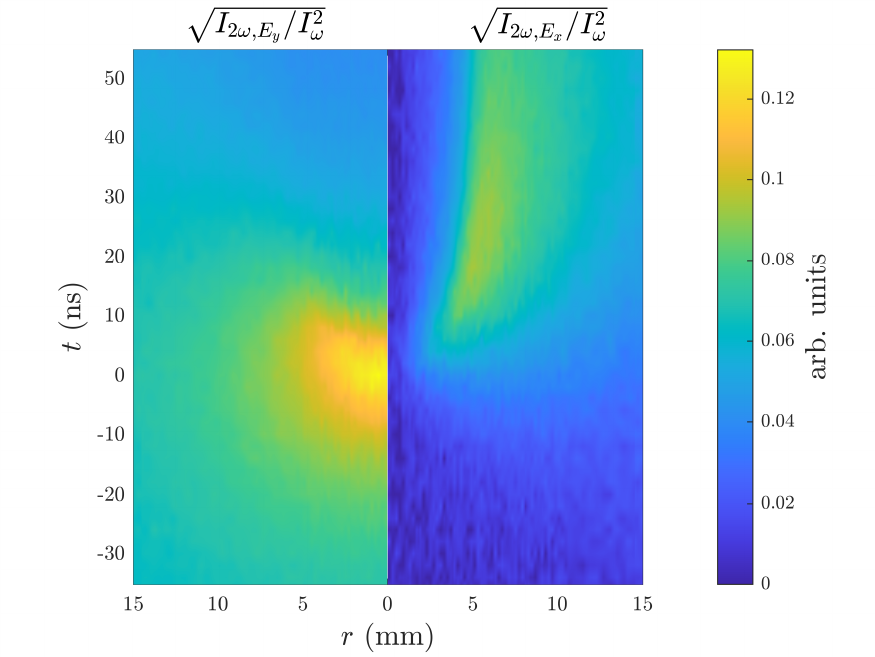}
  \caption{\small Measured E-FISH signals for $E_y$ (left) and $E_x$ (right) of a streamer in 70\,mbar air with an applied voltage of 9.5\,kV. \changedtext{$t=0$ is defined as the time when $E_y$ peaks.}}
  \label{fig:signalMeasured}
\end{figure}

Figure~\ref{fig:signalMeasured} shows the amplitudes of the E-FISH signals for $E_y$ and $E_x$ of a single-channel streamer in 70\,mbar air with an applied voltage of 9.5\,kV. 
It is worth noting that the distribution of the $E_x$ signals can become asymmetric around $r=0$ due to interference between the \efish{} signal and the background signal.
We discuss this issue, its implications on other \efish{} results and corresponding solutions in detail in the supplementary material~\cite{Supplementary}.

\begin{figure}[!htb]
  \centering
  \includegraphics[width=0.5\textwidth]{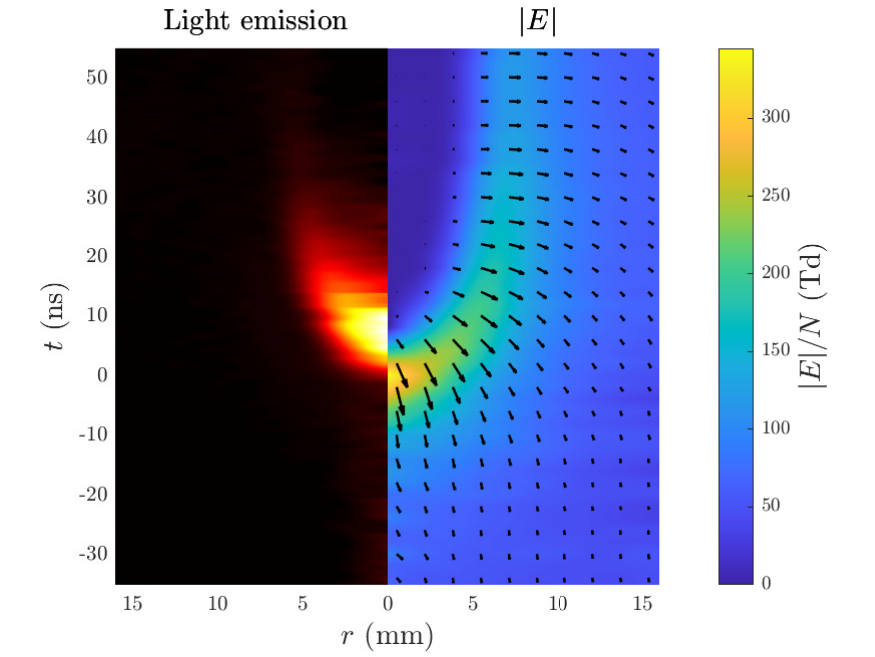}
  \includegraphics[width=0.5\textwidth]{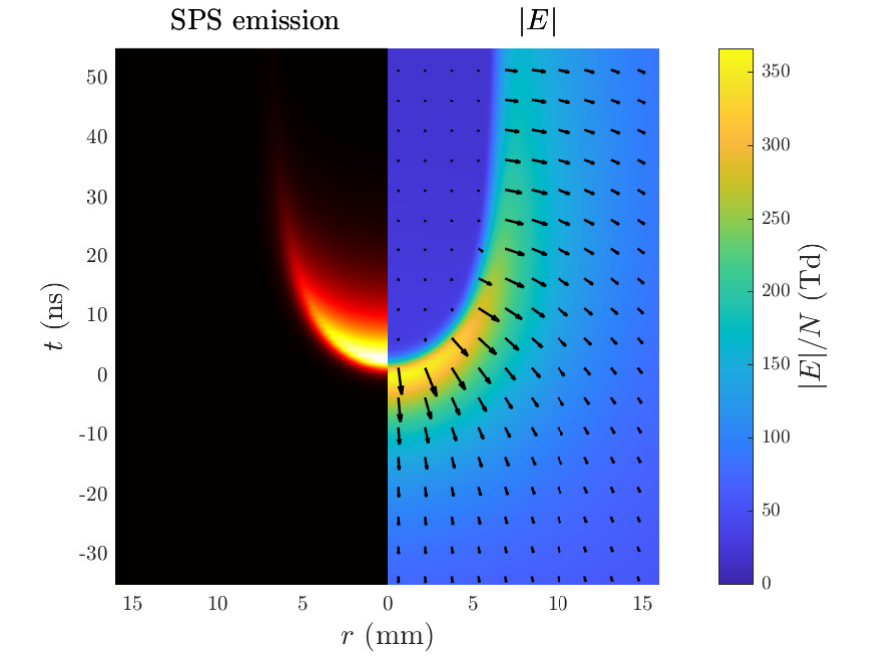}
  \caption{\small Experimental (top) and simulation (bottom) results of local \changedtext{normalized} light emission intensity and \changedtext{reduced} electric field distribution.}
  \label{fig:emissionAndField}
\end{figure}

By using the deconvolution method described, and multiplying with calibration constant $C_\mathrm{cal}$, the absolute values of both the $r$ and $y$ components of the electric field can be restored from the signals. 
The magnitude of the electric field is calculated as $\lvert E \rvert = \sqrt{E_r^2+E_y^2}$.
Figure~\ref{fig:emissionAndField} shows the comparison of local light emission intensity and electric field distribution between experiment and simulation results. \changedtext{$\lvert E \rvert/N$ is the absolute reduced electric field in Townsend (Td, $E/N$ in units $10^{-21}$\,V$\cdot$m$^2$), where $N$ is  the gas number density. Here, $N$ equals $1.88 \cdot 10^{24}$\,m$^{-3}$ at 70 mbar when assuming room temperature.}
In the simulation, the conditions in~\cite{Li2021} are adapted to fit the voltage and pressure used in this experiment. 
The emitted light is estimated by the $\mathrm{N_2(C^3\Pi_u)}$ density because the SPS transition is responsible for most of the optical emission under our discharge conditions~\cite{Pancheshnyi2000}.
Note that the simulation only treats one voltage pulse, therefore the heating effect of repetitive pulses is not taken into account. In the experiments the gas temperature will likely be elevated above room temperature~\cite{Adams2019,Pai2010}, but its exact value cannot be determined in this setup.
\\ \indent Qualitatively, the experimental and simulation results exhibit very high visual similarity.
In both cases the space charge layer with the crescent shape of the streamer head is clearly visible.
The electric field is most intense around the head of streamers and mostly pointing downwards.
Inside the streamer channel, the electric field almost vanishes, as the highly conductive streamer channel shields the electric field very efficiently. 
Furthermore, the light emission always lags behind the electric field~\cite{Wagenaars2007,Hoder2015}. 
Thus, light emission based methods to measure the electric field can only determine the field in the area behind the peak electric field, while E-FISH can resolve the field distribution completely.
In the simulation, the light emission profile shows a sharp front, while in the experiment it is smeared out more and shows some internal structure. The smearing is partly due to the camera integration time and streamer jitter, while the internal structure is attributed to oscillating ripples in the voltage waveform that change the velocity of the streamers.
\\ \indent Quantitatively, the streamers in both cases have similar size and very close field magnitude.
The electrodynamic diameter measured by the electric field is defined as the radial position at which $E_r$ reaches a maximum.
In experiment and simulation these are 7.4\,mm and 7.8\,mm respectively for $t=50\,$ns, which agrees very well.

\begin{figure}[!htb]
  \centering
  \includegraphics[width=0.5\textwidth]{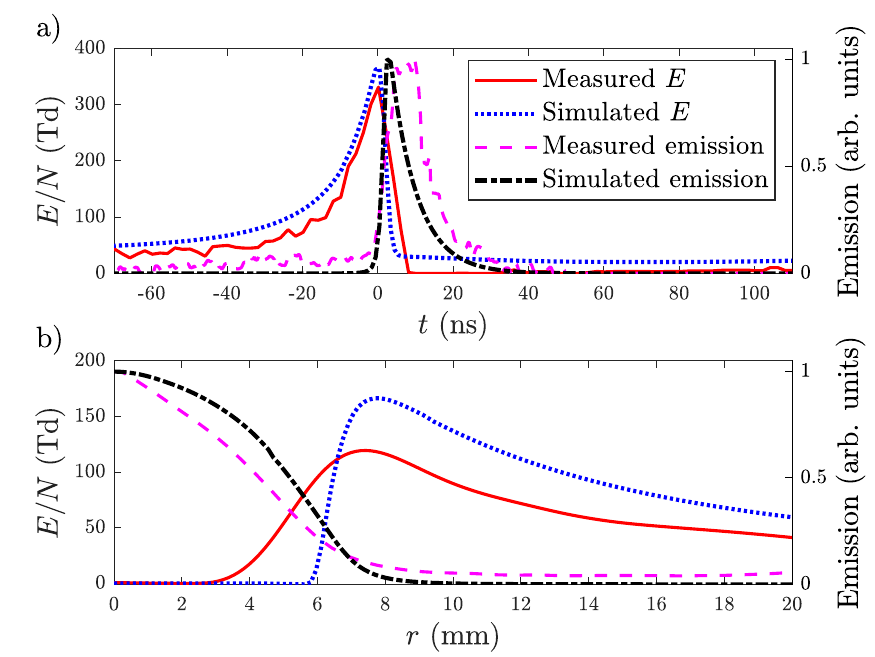}
  \caption{\small A comparison of \changedtext{a)} the axial field ($E_y$) and light emission and \changedtext{b)} the radial field ($E_r$) and light emission  between \efish{} experiment and simulation. The radial field was determined at 50\,ns while the radial light emission was integrated over the entire time domain and normalized to its maximum.}
\label{fig:axialField}
\end{figure}

Figure~\ref{fig:axialField} shows a comparison of the axial field evolution and light emission, and of the radial field and light emission at 50\,ns between experiment and simulation. 
In figure~\ref{fig:axialField}(a), it is shown that the simulated and measured electric field stay roughly constant at the background field level and rise rapidly at around $t=-20$\,ns when the streamer head is approaching the probing laser. The maximum electric field peaks at 364\,Td and 320\,Td for simulation and experiment, respectively. \changedtext{These values correspond to $6.4\cdot10^5$\,V/m and $5.6\cdot10^5$\,V/m respectively. The difference between experiment and simulation is likely caused by the temporal resolution of the experiment being limited by the streamer jitter and by a slightly elevated gas temperature in the experiments.}

Subsequently, the axial field drops drastically within 5\,ns and remains at almost zero, as the streamer leaves behind a conducting channel after the streamer head crosses. 
The peak value is higher than the breakdown threshold in air ($\sim$120\,Td) and is lower than the field determined by OES in~\cite{Dijcks2023a} with a peak field of 540\,Td in 33\,mbar air and in~\cite{Hoder2015} around 500\,Td at atmospheric pressure.
Mrkvi\u{c}kov\'{a} \textit{et al.}~\cite{Mrkvickova2023} compared the electric field in an atmospheric pressure Townsend discharge in nitrogen determined by \efish{} and OES, and found that the OES method gives systematically higher values.
They attributed this to the omission of additional population processes of $\mathrm{N_2^+(B^2\Sigma _u^+)}$.
In figure~\ref{fig:axialField}(b), the radial field profiles have similar shapes, except that the measured field has a smoother edge and lower peak value.
This is probably due to the lower field intensity of the calibration measurement as well as a larger background signal of $E_x$, which results in a lower sensitivity. 
Outside the streamer channel, the radial fields in both cases decay with approximately the same speed. 


In conclusion, we have used \efish{} to determine for the first time an experimentally obtained detailed direction-resolved spatiotemporal distribution of the electric field in a single-channel streamer. Moreover, the electric field in areas with little to no light emission is revealed.
The measurement was performed in 70\,mbar air with resolutions of 0.2\,mm and 2\,ns. Simultaneously, the optical emission was tracked with a resolution of 2 ns.
We developed a deconvolution method for \efish~including the effect of phase variations and designed dedicated calibration experiments for cylindrically symmetrical fields to restore the absolute electric field distribution from \efish{} signals. 
Simulations on the electric field and SPS emission of the same streamer were obtained using a 2D axisymmetric drift-diffusion-reaction fluid model. 
We compared the experimental and simulation results and they show great agreement both qualitatively and quantitatively.
The maximum electric field peaks at 364\,Td and 320\,Td for simulation and experiment, respectively.
This enables us to further verify and validate the simulation models and have a comprehensive understanding of streamer discharges in nearly all gas mixtures, both on the development and on the chemical processes inside. 

\changedtext{It must be noted that for our discharge geometry and
laser set-up, the difference in outcome between our full processing
method and a more standard Abel inversion, for $E_y$ is below 20\%.
However, the standard Abel inversion is more sensitive to noise
from the outer regions and therefore requires more smoothing.
For $E_r$, a standard Abel inversion is
insufficient, as it cannot translate the $E_x$ vector into $E_r$.}

The developed method for analysing \efish~measurements can be applied on highly repetitive plasmas/electric fields as a scan of the electric field is required. However, cylindrical symmetry is needed. 
\\ \indent Future efforts should be directed to adjusting the analysis and corresponding methods to make \efish~suitable for asymmetric fields. 
This will allow for high resolution direct electric field measurements of transient/fast moving electric fields in (ionized) gases, which is currently impossible.

Y G was supported by the China Scholarship Council (CSC) Grant No. 202006280041.

%

\end{document}


\title{Supplementary material:\\
Measurement of the electric field distribution in streamer discharges}
\author{Yihao Guo$^1$, Anne Limburg$^1$, Jesse Laarman$^1$, Jannis Teunissen$^2$ and Sander Nijdam$^1$}%
\affil{\textit{$^1$Department of Applied Physics and Science Education, Eindhoven University of Technology, 5600 MB Eindhoven, The Netherlands}\\ \textit{$^2$Centrum Wiskunde \& Informatica (CWI), 1090 GB Amsterdam, The Netherlands}}
\maketitle

\section{Signal inversion}
\label{sec:inversion}

As explained in the main text, we measure a signal $s(x,t)$ that results from either the longitudinal field $E_y$ of an axisymetric streamer or from its the radial field $E_r$.
These cases correspond to the following integrals

\begin{align}
  s_y(x,t) &= C_\mathrm{cal} \left| \int_{-L}^L  g(z) E_y(r,t)  \, \mathrm{d}z \right |,\label{eq:sx-Ey}\\
  s_r(x,t) &= C_\mathrm{cal} \left| \int_{-L}^L  g(z) \, x/r \, E_r(r,t)  \, \mathrm{d}z \right |,\label{eq:sx-Er}
\end{align}
where $r = \sqrt{x^2 + z^2}$ and
\begin{equation}
  \label{eq:integral-y-v2}
  g(z) = \frac{e^{-i\Delta k z}}{1 + i \frac{z}{z_R}}.
\end{equation}
The meaning and values of $\Delta k$ and $z_R$ are discussed in the main text.
Below, we will for brevity refer to $E_y(r,t)$ or $E_r(r,t)$ simply as $f(r)$. The dependence on $t$ is left out as the inversion is performed for every $t$ (line-by-line).
The goal is to approximate $f(r_i)$ at given locations $r_i$ ($i = 1, \dots, N$) from a set of samples $s(x_j)$ ($j = 1, \dots, M$), which is an inverse problem.
If we have samples $\mathbf{f} = (f_1, f_2, \dots, f_N)$, then we can approximate the integral for a given $x$ by weighted sum
\begin{equation}
  s(x_j) \approx \left| \sum_{i = 1}^{N} w_i f(r_i) \right |,
\end{equation}
where the weights $w_i$ still have to be determined, and where the factor $C_\mathrm{cal}$ has been left out for brevity.
When we consider multiple $x$-positions simultaneously, we can express these sums as an approximately linear system of the form
\begin{equation}
  \label{eq:discrete-system}
  |\mathbf{W} \mathbf{f}| = \mathbf{s},
\end{equation}
where $\mathbf{f}$ and $\mathbf{s}$ are real vectors, $\mathbf{W}$ is a complex matrix, and the norm operator $||$ is applied element wise.
Note that for any solution $\mathbf{f}$, $-\mathbf{f}$ is also a solution.
We can solve such a system in the least-squares sense by minimizing the expression
\begin{equation}
  |\left(|\mathbf{W} \mathbf{f}| - \mathbf{s}\right)|.
\end{equation}
In practice, it is typically necessary to include some regularization, to ensure that a unique solution exists and that the solution is not too sensitive to noise in the measurements.
We include such regularization by minimizing the following expression
\begin{equation}
  \label{eq:discrete-system-reg}
  |\left(|\mathbf{W} \mathbf{f}| - \mathbf{s}\right)| + c \, |\mathbf{f}''|,
\end{equation}
where $\mathbf{f}''$ is an approximation of the second derivative of $\mathbf{f}$, given by $f_i'' = (f_{i-1} - 2 f_i + f_{i+1})/(\Delta r)^2$, with $\Delta r$ being the spacing between sampling points.
For the inversions shown in the paper, we used $c = 3 \times 10^{-2} \, (\Delta r)^2$.


\subsection{Obtaining the weight matrix}

Below, we explain how we obtain the weight matrix $\mathbf{W}$ in equation (\ref{eq:discrete-system}).
The first step is to convert the line integrals into integrals over the radial coordinate $r$, as is also done in a standard Abel transform.
Suppose we want to evaluate the following integral:
\begin{equation}
  s(x) = \int_{-\infty}^{\infty} f(\sqrt{x^2 + z^2}) g(z) \mathrm{d}z,
\end{equation}
where $r = \sqrt{x^2 + z^2}$, as above.
We can change the integration variable to $r$, noting that
\begin{equation}
  \label{eq:var-change}
  z = \pm \sqrt{r^2 - x^2}, \quad \mathrm{d}z = \pm \frac{r \mathrm{d}r}{\sqrt{r^2 - x^2}}.
\end{equation}
Now split the integral in two parts
\begin{equation}
  s(x) = \int_{-\infty}^{0} f(\sqrt{x^2 + z^2}) g(z) \mathrm{d}z + \int_{0}^{\infty} f(\sqrt{x^2 + z^2}) g(z) \mathrm{d}z,
\end{equation}
and then change variable in both parts, using the appropriate signs from equation~\eqref{eq:var-change}:
\begin{equation}
  s(x) = \int_{\infty}^{|x|} f(r) g(-\sqrt{r^2 - x^2}) \frac{-r}{\sqrt{r^2 - x^2}}\mathrm{d}r + \int_{|x|}^{\infty} f(r) g(\sqrt{r^2 - x^2}) \frac{r}{\sqrt{r^2 - x^2}} \mathrm{d}r.
\end{equation}
This can be simplified to
\begin{equation}
  \label{eq:transform-full}
  s(x) = \int_{|x|}^{\infty} f(r) \left[g(\sqrt{r^2 - x^2}) + g(-\sqrt{r^2 - x^2})\right] \frac{r}{\sqrt{r^2 - x^2}} \mathrm{d}r.
\end{equation}
To numerically evaluate equation \eqref{eq:transform-full}, we for simplicity assume that the $\mathbf{f} = (f_1, f_2, \dots, f_N)$ are equally spaced, so that $r_{i+1} - r_i = \Delta r$.
We furthermore assume that $f(r)$ and $g(z)$ do not vary strongly in each interval from $r_i - 0.5 \Delta r$ to $r_i + 0.5 \Delta r$.
On the other hand, the factor $\frac{r}{\sqrt{r^2 - x^2}}$ will vary significantly (and diverge) for $r \to |x|$.
Since
\begin{equation}
  \int \frac{r}{\sqrt{r^2 - x^2}} \mathrm{d}r = \sqrt{r^2 - x^2} + C,
\end{equation}
we can handle this by analytically integrating this factor
\begin{equation}
  \int_{r_i - 0.5 \Delta r}^{r_i + 0.5 \Delta r} \frac{r}{\sqrt{r^2 - x^2}} \mathrm{d}r = \sqrt{(r_i + 0.5 \Delta r)^2 - x^2} - \sqrt{(r_i - 0.5 \Delta r)^2 - x^2},
\end{equation}
and then changing the integral to a sum:
\begin{equation}
  \label{eq:transform-sum-Ey}
  s(x) \approx \sum_{i = 1}^{N} f(r_i) \left[g(\sqrt{r_i^2 - x^2}) + g(-\sqrt{r_i^2 - x^2})\right] \left( \sqrt{(r_i + 0.5 \Delta r)^2 - x^2} - \sqrt{(r_i - 0.5 \Delta r)^2 - x^2} \right).
\end{equation}
Since the integral in equation \eqref{eq:transform-full} starts from $r = |x|$, only terms in the sum for which $r^2 - x^2 > 0$ should contribute.
This can be achieved by replacing all terms of the form $r^2 - x^2$ by zero when they are negative.

If we consider equation (\ref{eq:sx-Er}) for $E_r$, there will be an extra factor $\sin(\theta) = x/r$ in equation \eqref{eq:transform-full}, so that
\begin{equation}
  \label{eq:integral-x}
  s(x) = \int_{|x|}^{\infty} f(r) \left[g(\sqrt{r^2 - x^2}) + g(-\sqrt{r^2 - x^2})\right] \frac{x}{\sqrt{r^2 - x^2}} \mathrm{d}r.
\end{equation}
The weight function for each sample $f_i$ can again be integrated analytically:
\begin{equation}
  \int \frac{x}{\sqrt{r^2 - x^2}} \mathrm{d}r = x \log\left(2 \sqrt{r^2 - x^2} + 2 r \right) + C.
\end{equation}
This results in a sum
\begin{equation}
  \label{eq:transform-sum-Ex}
  s(x) \approx \sum_{i = 1}^{N} f(r_i) \left[g(\sqrt{r_i^2 - x^2}) + g(-\sqrt{r_i^2 - x^2})\right] \, x \,
  \log\left(\frac{\sqrt{r_b^2 - x^2} + r_b}{\sqrt{r_a^2 - x^2} + r_a} \right),
\end{equation}
where $r_a = r_i - 0.5 \Delta r$ and $r_b = r_i + 0.5 \Delta r$.
Note that we used $\log(b) - \log(a) = \log(b/a)$.
Again, the integral in equation~\eqref{eq:integral-x} starts from $r = |x|$, so the sum should only contain corresponding contributions.
This means that $\sqrt{r_a^2 - x^2} + r_a \geq |x|$ and similarly for $r_b$, so that these terms can be replaced by $\max(\sqrt{r_a^2 - x^2} + r_a, |x|)$.
Since $\lim_{x \to 0} x \log(x) = 0$, the case $x = 0$ should give a zero contribution.

The right-hand sides of equations (\ref{eq:transform-sum-Ey}) and (\ref{eq:transform-sum-Ex}) contain the weights $w_i$ for each of the $f(r_i)$.
Computing these weights for different sampling locations $x_j$ (i.e., where $s(x)$ is measured) results in the full weight matrix $\mathbf{W}$ used in equation (\ref{eq:discrete-system}).

\section{Calibration method}\label{sec:calibration}

To obtain the absolute value of the electric field, a calibration measurement on a known field is performed.
Chng~\cite{Chng2020} \textit{et al.} found that the measured E-FISH signal is strongly influenced by the electric field profile due to the Gouy phase shift of the focused beam.
Therefore, the conventional calibration method for E-FISH experiments using Laplacian fields can lead to large errors. 
Nakamura~\cite{Nakamura2022} \textit{et al.} evaluated the dependence of the error on the Rayleigh length and the targeted field length when applying the conventional calibration method.
To avoid these errors, the calibrated field should have a similar spatial profile as the targeted electric field.

\begin{figure}
  \begin{subfigure}[b]{0.5\textwidth}
      \centering
    \includegraphics[width=7cm]{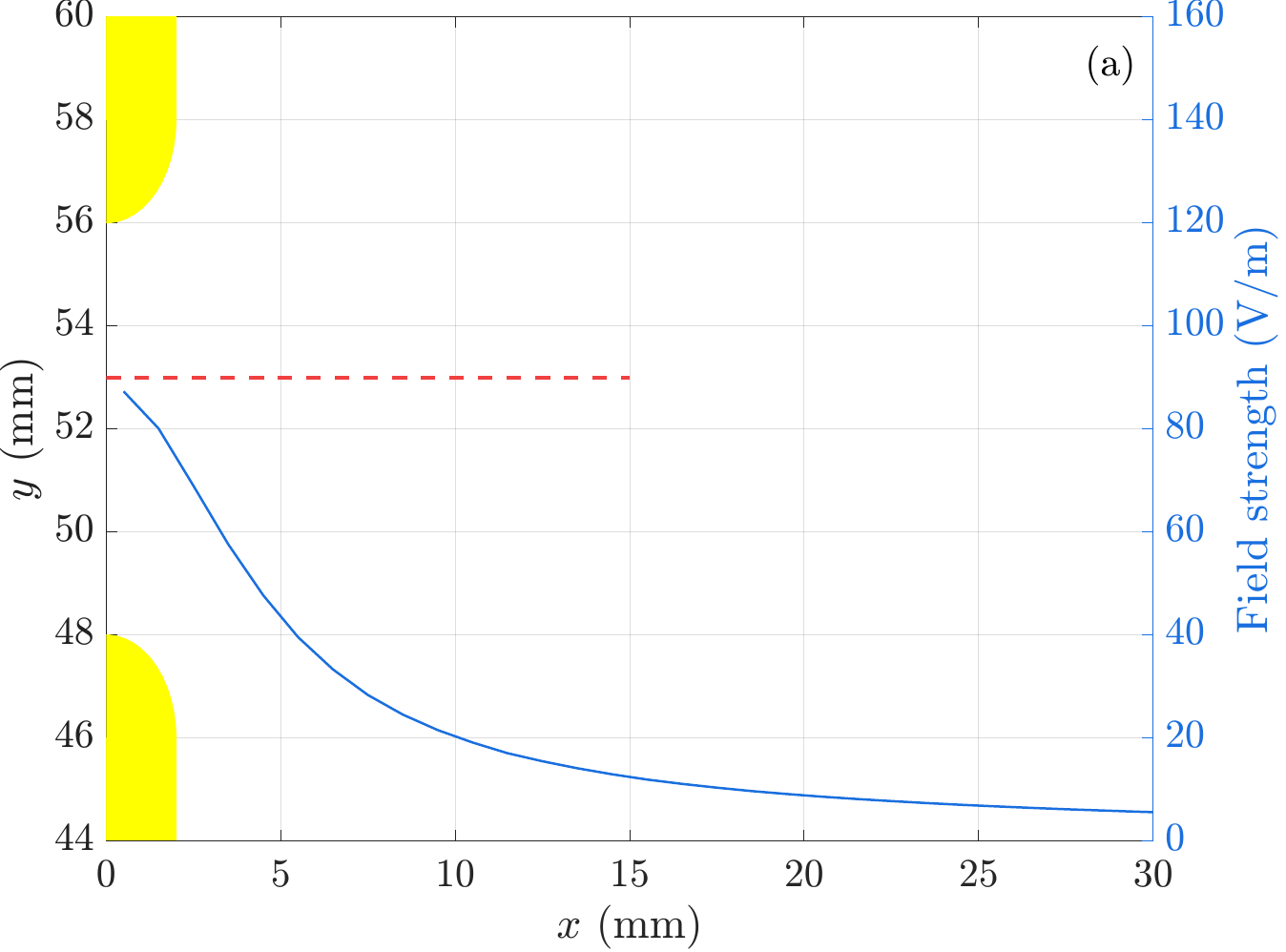}
    \caption{}
  \end{subfigure}
  \begin{subfigure}[b]{0.5\textwidth}
      \centering
    \includegraphics[width=7cm]{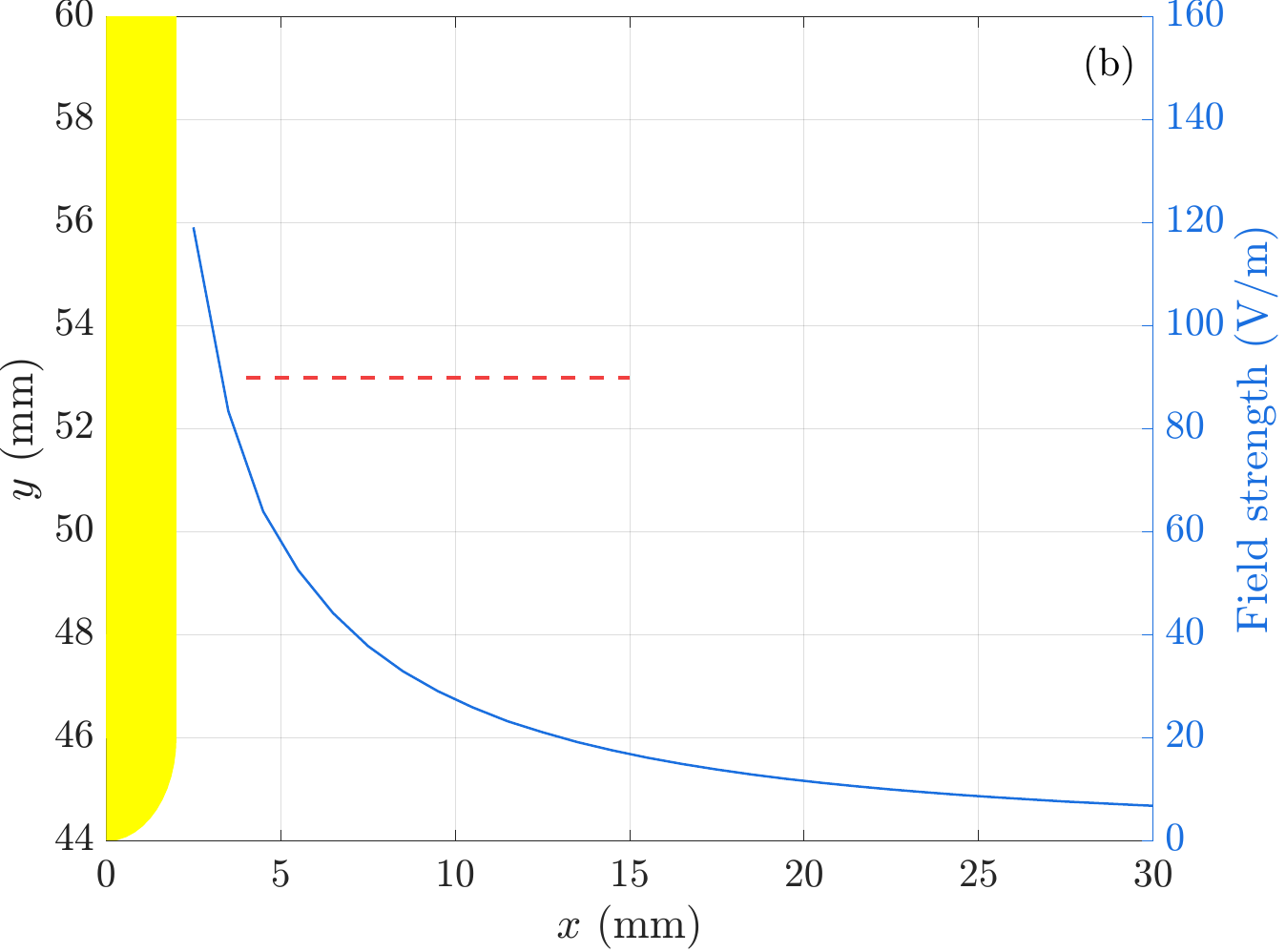}
    \caption{}
  \end{subfigure}
  \caption{\small Calculated fields of (a) rod-to-rod electrodes for $E_y$ and (b) rod-to-cylinder electrodes for $E_x$, respectively. The applied voltage for both configurations is 1\,V. The shapes colored in yellow show the electrode geometry. The dotted red line indicates the height and the scan range of the laser.}
  \label{fig:calibrationSim}
\end{figure}

In this work, we designed rod-to-rod and rod-to-cylinder electrodes for the calibration of $E_y$ and $E_x$, respectively.
Figure~\ref{fig:calibrationSim} shows the calculated fields of the two pairs of calibration electrodes simulated by COMSOL Multiphysics. 
The field profiles have similar size and shape as the streamer generated under the conditions used in this work.
The calculated fields are then forward-transformed to become the ``calculated E-FISH signals''.

We then measure the E-FISH signals of the calibration electrodes.
To increase the signal intensity, the highest voltages for both configurations below the breakdown threshold are used, which are 1.5\,kV and 2.5\,kV for $E_y$ and $E_x$, respectively.
Because the measurement range is limited within $\pm$15\,mm due to the size of the window on the vessel, the measured signals are much shorter than the simulated ones and do not reach 0 at the boundary.
The calibration constant $C_{\mathrm{cal}}$ is obtained by comparing the calculated and the measured E-FISH signals on a range of $r$ over which both signals are reliable.
For $E_y$ this is $r = 2 - 6$\,mm and for $E_x$ this is $r = 5 - 10$\,mm.
This method shows robust results and little dependence on the exact parameters used.

\section{Full processing method}

In order to convert the measured intensities to electric field strengths, the following procedure is followed:
\begin{enumerate}
    \item The intensities of the SH $I_{2\omega}$ and the fundamental beam $I_{\omega}$ are obtained by integrating the peak of the recorded traces over 4\,ns. Then the measured signal per point is calculated as $s_{meas}=\sqrt{I_{2\omega}/I_{\omega}^2}$. To increase the signal-to-noise ratio, 40 measurements are averaged.
    \item The background signal is subtracted using $s(x) = \sqrt{s_{meas}(x)^2 - s_{BG}^2}$, where $s_{meas}(x)$ is the measured signal as discussed above, and $s_{BG}$ is the background signal, probably due to interference as discussed in section \ref{sec: interference}. The background is determined by a separate measurement without field, or, for $E_x$ using the area below the streamer where there is no field expected. In this case the background is a function of the $x$ coordinate.
    \item The $s(x)$ data is made symmetric by flipping one side (in the $x$ direction, around \changedtext{the symmetry axis}) and averaging it with the other side.
    \item The backwards inversion discussed in section~\ref{sec:inversion} is used to convert the signal. The entire domain up to $r=150$\,mm (edge of the vessel) is used in this inversion. The solution is forced to be fully positive.
    \item The solution is multiplied with $C_{\mathrm{cal}}$ as discussed in section~\ref{sec:calibration} and converted to Townsend using the temperature and pressure.
\end{enumerate}

\section{Interference with other signals.} 
\label{sec: interference}
Second harmonic (SH) generation can happen through E-FISH, but also through asymmetries in solid materials. The latter is intentionally done in doubling crystals, which are used in lasers. In E-FISH measurements, generally lenses, mirrors, prisms and other optics are used as part of the setup. These components can generate unwanted SH radiation which is colinear and coherent with the E-FISH signal. In most works, this background signal is minimized by lowering the power of the input beam. The left-over background signal is then subtracted from the E-FISH measurements. However, the background SH can interfere with the E-FISH signal, which results in the following relation:
\begin{equation}
    I = I_E+I_B+2 \sqrt{I_E I_B} \cos{\phi},
\end{equation}
with $I$ the total intensity measured by the SH detector, $I_B$ the background SH intensity, $I_E$ the E-FISH signal and $\phi$ the phase difference between the background SH and E-FISH signal. The latter is proportional to the distance between the E-FISH measurement area and the SH-generating optics and to the wavevector mismatch (and thereby the gas type and density).

$\cos{\phi}$ switches sign for changing electric field polarity, which results in 
\begin{equation}\label{eq:Intensity_interference_two_versions}
\begin{split}
    I_+ &= I_E+I_B+2 \sqrt{I_E I_B} \cos{\phi},\\
    I_- &= I_E+I_B-2 \sqrt{I_E I_B} \cos{\phi},    
\end{split}
\end{equation}
where the two different components result from opposite field directions in the E-FISH measurement area (and thereby opposite phase at the point of interference).

Due to the asymmetry in equation~\eqref{eq:Intensity_interference_two_versions}, measured signals can be higher or lower than the real E-FISH signals. Simply subtracting the background ($I_B$) could therefore lead to negative measured intensities, which is clearly nonphysical. Also, when measuring electric fields which have two different polarities, which are expected to be symmetric either in space or time, this can lead to asymmetric measurement results.

\begin{figure}[!htb]
  \centering
  \includegraphics[width=0.5\textwidth]{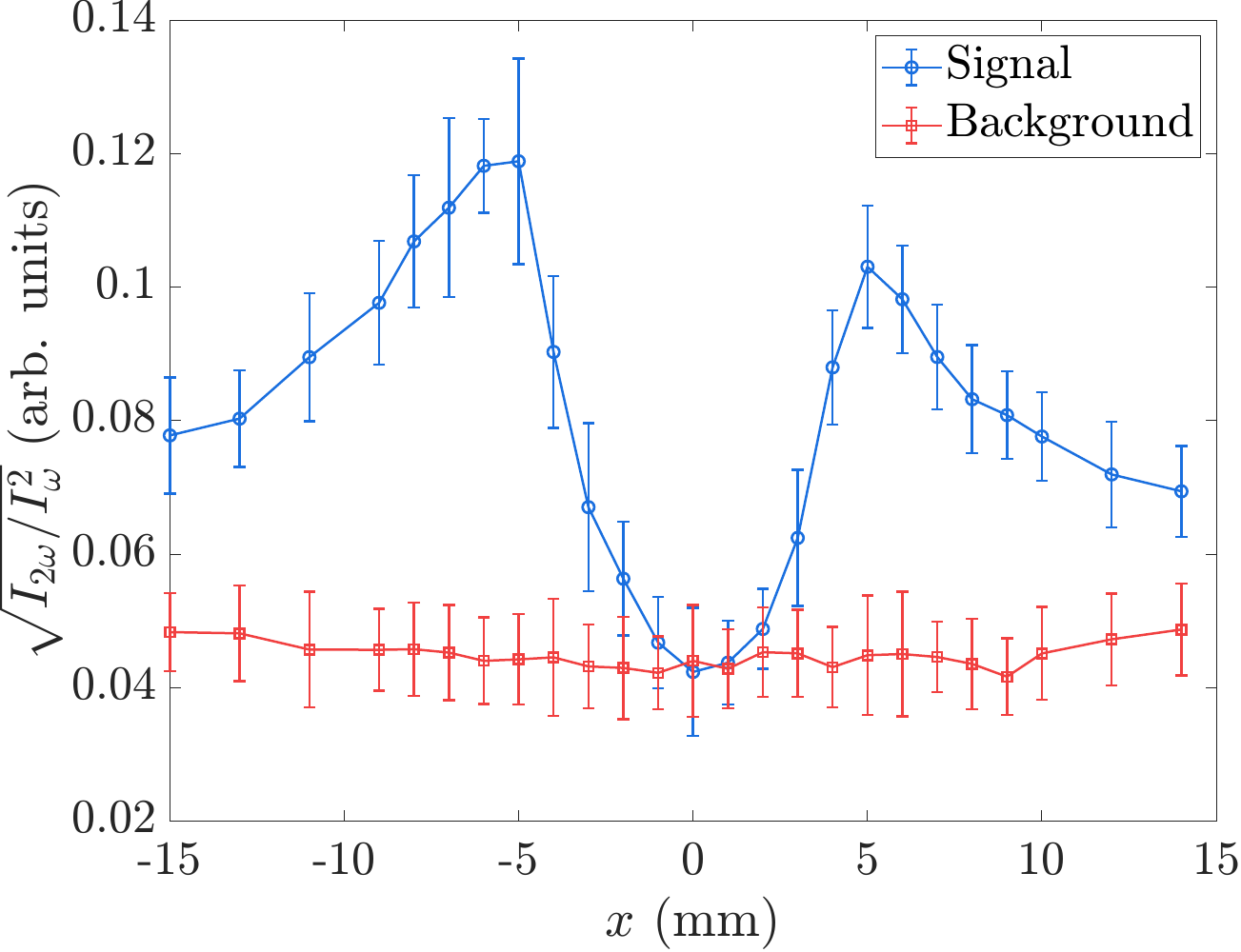}
  \caption{\small An example of asymmetric $E_x$ measurement on a propagating single-channel streamer and the corresponding background signals.}
  \label{fig:asymmetricEx}
\end{figure}

Examples of this are phase-resolved measurements of (Laplacian) electric fields generated by applying a sinusoidal voltage to symmetric electrodes, leading to an asymmetry between the measured fields during the two half-waves. This has been observed and discussed by~\cite{Mrkvickova2023}.
In our experiments, the electric field in the $x$-direction is expected to be symmetric around the streamer propagation axis due to the cylindrical symmetry. There, the field direction is reversed between the left and right halves. The asymmetry in these measurements can be seen in  figure~\ref{fig:asymmetricEx}.

\begin{figure}
  \begin{subfigure}[b]{0.5\textwidth}
      \centering
    \includegraphics[width=7cm]{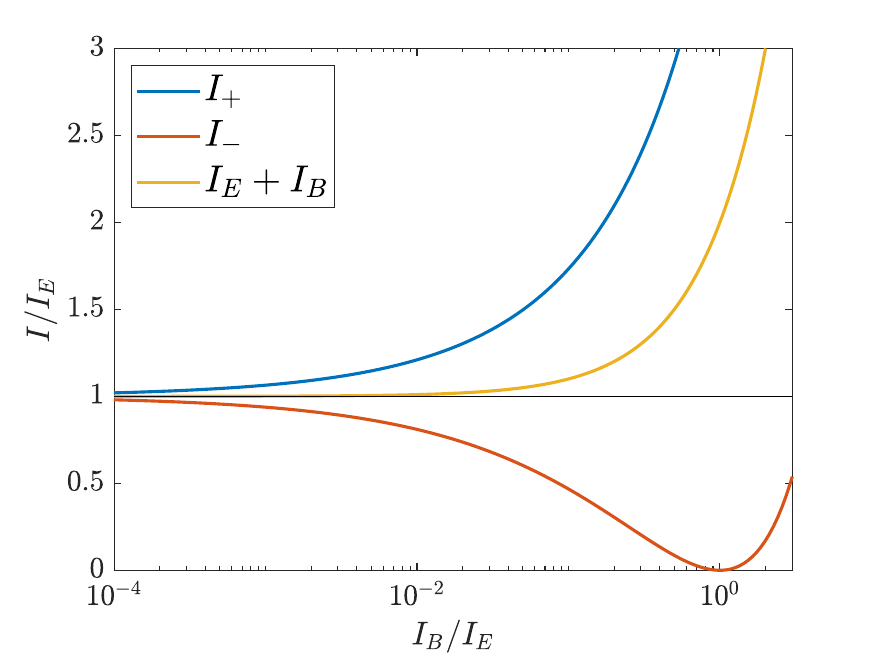}
    \caption{}
  \end{subfigure}
  \begin{subfigure}[b]{0.5\textwidth}
      \centering
    \includegraphics[width=7cm]{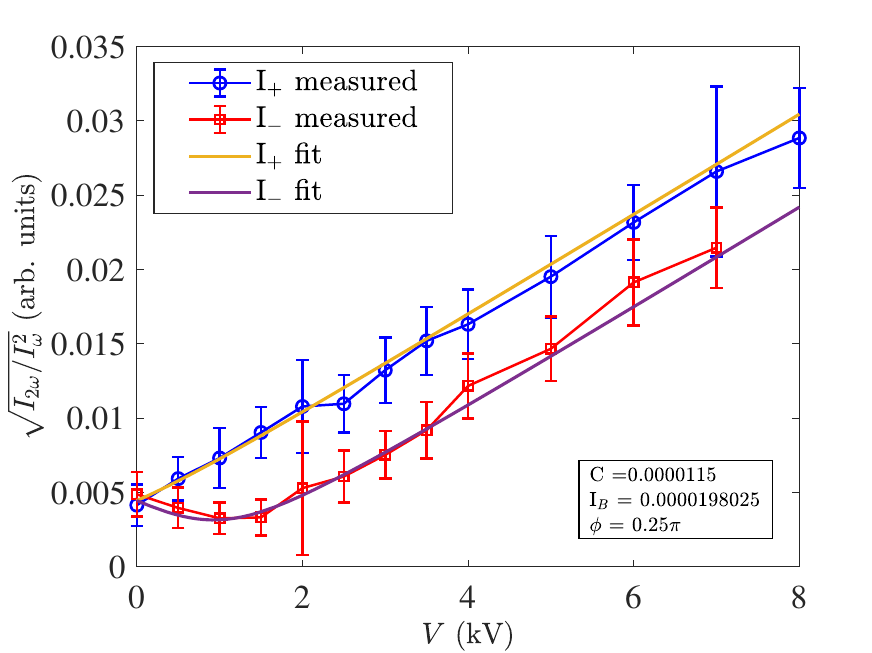}
    \caption{}
  \end{subfigure}
  \caption{\small (a) Calculated ratio between the expected SH intensities for both field polarities ($I_+$ and $I_-$) and the E-FISH intensity, or the naively expected intensity ($I_E$ + $I_B$) and the E-FISH signal as function of relative background level for $\cos{\phi}=1$.
  (b) Measured E-FISH signal between symmetric electrodes with field in the $x$-direction under opposite polarities. Includes a fit of the calibration constant, the background intensity and $\cos{\phi}$.}
  \label{fig:signalsCompared}
\end{figure}

The error in the results can be much higher than the background signal itself due to the $2 \sqrt{I_E I_B}$ term; this is illustrated in figure~\ref{fig:signalsCompared}.

For measurements which do not have symmetric dual polarity signals, it is very difficult to determine if $\cos{\phi}$ is close to zero, which would allow a simple background subtraction.

There are a few ways to deal with the issue discussed above:
\begin{enumerate}
    \item Change the distance between E-FISH measurement area and SH-producing optics such that $\cos{\phi}\approx0$ and a simple background subtraction is allowed. However, changes in ambient air pressure can already lead to deviations from this ideal case. For example, the horizontal distance between the discharge gap and the prism is about 1\,m in our setup. A change of 20\,mbar in air pressure changes cos $\phi$ to 0.3.
    \item For purely anti-symmetric cases (like $E_x$ in our experiments), averaging both polarities (sides) suffices because in this case the last term in equations~\eqref{eq:Intensity_interference_two_versions} drops out.
    \item Minimizing the production of SH by windows and optics by lowering beam intensity and carefully choosing and cleaning the optics. One way to do it is using cross-beam E-FISH~\cite{Raskar2022}, where no beam separation optics is required.
\end{enumerate}
If none of these are (or can be) applied, the background SH signal can lead to large errors in the measured E-FISH signals and thereby in the determined electric fields.

 

\bibliographystyle{unsrt}
